\def\z0{\rm Z^0}
\newcommand{\as}{\alpha_{\rm s}}

\newcommand{\oaa}{{\cal O}(\as^2)}
\newcommand{\oaaa}{{\cal O}(\as^3)}
\newcommand{\epem}{\rm e^+\rm e^-}
\newcommand{\yc}{y_{\rm cut}}
\newcommand{\amz}{\as(M_{\rm Z^0})}
\newcommand{\oabel}{{\cal O}(\alpha_A^2)}
\def\mz{M_{\rm Z^0}}

\def\d2{D_2}
\def\oq{\char'134}
\def\lamsb{\Lambda_{\overline{MS}}}
\def\ecm{E_{cm}}

\def\m2{\mu^2}
\def\q{\rm q}

\def\q2{Q^2}
\def\asq{\as (\q2 )}
\documentstyle[epsf,twoside,fleqn,espcrc2]{article}

\begin{document}
 
\title{EXPERIMENTAL TESTS OF ASYMPTOTIC FREEDOM}
\author{Siegfried Bethke\address{III. Physikalisches Institut, RWTH, D - 52056
Aachen, Germany}%
 }

\begin{abstract}
Measurements which probe the energy dependence of $\as$, the coupling strength
of the strong interaction, are reviewed.
Jet counting in $\epem$ annihilation, combining results obtained in the centre
of mass energy range from 22 to 133~GeV, provides direct evidence for an
asymptotically free coupling, without the need to determine explicit
values of $\as$. 
Recent results from jet production in $e\ p$ and in $p\
\overline{p}$ collisions, obtained in single experiments spanning large ranges of
momentum transfer, $\q2$, are in good agreement with the running of $\as$ as
predicted by QCD. 
Mass spectra of hadronic decays of $\tau$-leptons are analysed to
probe the running $\as$ in the very low energy domain, $0.7\
{\rm GeV}^2 < \q2 < M_{\tau}^2$.
An update of the world summary of measurements of $\asq$ consistently
proves the energy dependence of $\as$ and results in a combined average of
$\amz = 0.118 \pm 0.006$.
%
\vskip-80mm
   {\small \noindent
   Talk presented at the {\it QCD Euroconference 96}, Montpellier (France)
July 4-12, 1996.} 
   \begin{flushright} {\large PITHA 96/30} \\
    {\large September 1996}
   \end{flushright}
\vskip65mm 
\end{abstract}
 
\maketitle
 
\section{INTRODUCTION}

Since Quantum Chromodynamics \cite{qcd} and the concept of
asymptotic freedom \cite{asymf} were introduced to describe the dynamics of
hadronic processes at high momentum transfers, 
several \oq key" predictions of the theory were successfully tested
by experiment:
`Evidence for jet structure in hadron production by $\epem$ annihilation' was
found in 1975 \cite{hanson}, the gluon was explicitly observed at the PETRA
$\epem$ storage ring in 1979 \cite{gluon}.
The first measurement of the coupling strength $\as$, the basic free parameter
of the theory, was reported in that same year \cite{as-mkj}, based on leading
order perturbative QCD.
The first determination of $\as$ in next-to-leading order (NLO) QCD dates back
to 1982 \cite{as-jade}.

After these pioneering years, many further tests of QCD were performed in $\epem$
annihilation, in deep inelastic lepton-nucleon scattering and at hadron
colliders.
In 1988, first evidence for the running $\as$ was
obtained from the energy dependence of 3-jet event production rates in $\epem$
annihilation
\cite{jadejet2}.
An update and a summary of these measurements will be presented in Section~3
of this review.

Although many determinations of $\as$,  in the
energy range of
$Q \sim 4$~to~$46$~GeV, were available by 1990, the
$running$ of $\as$ could not convincingly be seen from those results
\cite{altarelli}. 
Only in 1992, a compilation of measurements
of $\as$ in the energy range from 1.78~GeV (the mass of the
$\tau$-lepton) to 91.2~GeV (the mass of the $\z0$-boson), could
demonstrate the characteristic energy dependence of the strong coupling
\cite{sb-catani}.

The actual evidence for the running coupling strength or,
equivalently, for asymptotic freedom is summarised in this review.
The results from jet production rates in $\epem$
annihilation are presented in Section~3.
Recent studies of jet production and of the proton structure function $F_2$ in
deep inelastic electron-proton collisions are discussed in Section~4.
New results from jets in $p \overline{p}$ collisions and from hadronic decays
of $\tau$-leptons, demonstrating the energy dependence of $\as$ in ranges of
very high and very low momentum transfers, respectively, are presented in
Sections~5 and~6. 
An update of the world summary of $\as$ measurements
is finally given in Section~7.

This report is restricted to results which were published at the time
of this conference; 
preliminary results are
not taken into account.

\section{QCD  AND THE RUNNING $\as$}

Within perturbative QCD, the energy
dependence of $\as$ is given by the
$\beta$-function:
\begin{eqnarray}
\mu \frac{\partial  \as}{\partial \mu} & = & - \frac{\beta_0}{2\pi}
\as^2 - \frac{\beta_1}{4 \pi^2} \as^3 - \frac{\beta_2}{64\pi^3} \as^4
- \dots \nonumber \\
\beta_0 & = & 11 - \frac{2}{3} N_f\ , \nonumber \\
\beta_1 & = & 51 - \frac{19}{3} N_f \nonumber \\
\beta_2 & = & 2857 - \frac{5033}{9} N_f + \frac{325}{27} N_f^2\ ,
\end{eqnarray}
where $N_f$ is the number of quark flavours with
masses less than the energy scale $\mu$.
A solution of
Equation 1, in third order expansion, is \cite{pdg96} 
\begin{eqnarray}
\alpha_s(\mu) & = & \frac{4\pi}{\beta_0
\ln(\mu / \Lambda)^2} \big[
1 - 2\  \frac{\beta_1}{\beta_0^2}\
\frac{\ln\left(\ln( \mu / \Lambda)^2\right)}{\ln(\mu / \Lambda)^2}
\nonumber \\
& + & \frac{4
\beta_1^2}{\beta_0^4 \ln^2 (\mu / \Lambda)^2}
\big( \big( \ln \left[\ln (\mu / \Lambda)^2\right] -
\frac{1}{2}\big)^2 \nonumber \\
&+& \frac{ \beta_2 \beta_0}{8 \beta_1^2} - \frac{5}{4} \big) \big]~.
\end{eqnarray}

At large energy scales $\mu$, or equivalently at small distances,
$\as$ vanishes logarithmically; this behaviour of $\as$ is called
`asymptotic freedom'.

In this report all calculations, equations and results refer to
the `modified minimal subtraction scheme' ($\overline {\rm MS}$) \cite{msbar}.
More detailed information about the basic QCD equations and for the
treatment of heavy quark flavour thresholds can be found e.g. in
\cite{pdg96,msbar,wernerb,qcd94}.

\section{JET RATES IN $\epem$ ANNIHILATION}

Studies of hadron jets provide the most intuitive tests of the
underlying parton (i.e. quark and gluon) structure of hadronic events.
The most commonly used algorithm to reconstruct jets in $\epem$ annihilation was
introduced by the JADE collaboration
\cite{jadejet2}: the scaled pair mass of any two resolvable 
jets $i$ and $j$ in a hadronic event,  $y_{ij} = M_{ij}^2 /
E_{\rm vis}^2$, is required to exceed a  threshold value $\yc$, where
$E_{\rm vis}$ is the total visible (measured) energy of the event.  
In a recursive process, the pair of particles or clusters of particles
which has the smallest value of $y_{ij}$ is replaced by (or
`recombined' into) a single jet $k$ with four-momentum $p_k = p_i
+ p_j$, as long as $y_{ij} < \yc$.
The procedure is repeated until 
all $y_{ij}$ are larger than the jet resolution parameter $\yc$, 
and the remaining clusters of particles are
called jets.

\begin{figure}[htb]
\epsfxsize7.5cm
\epsffile{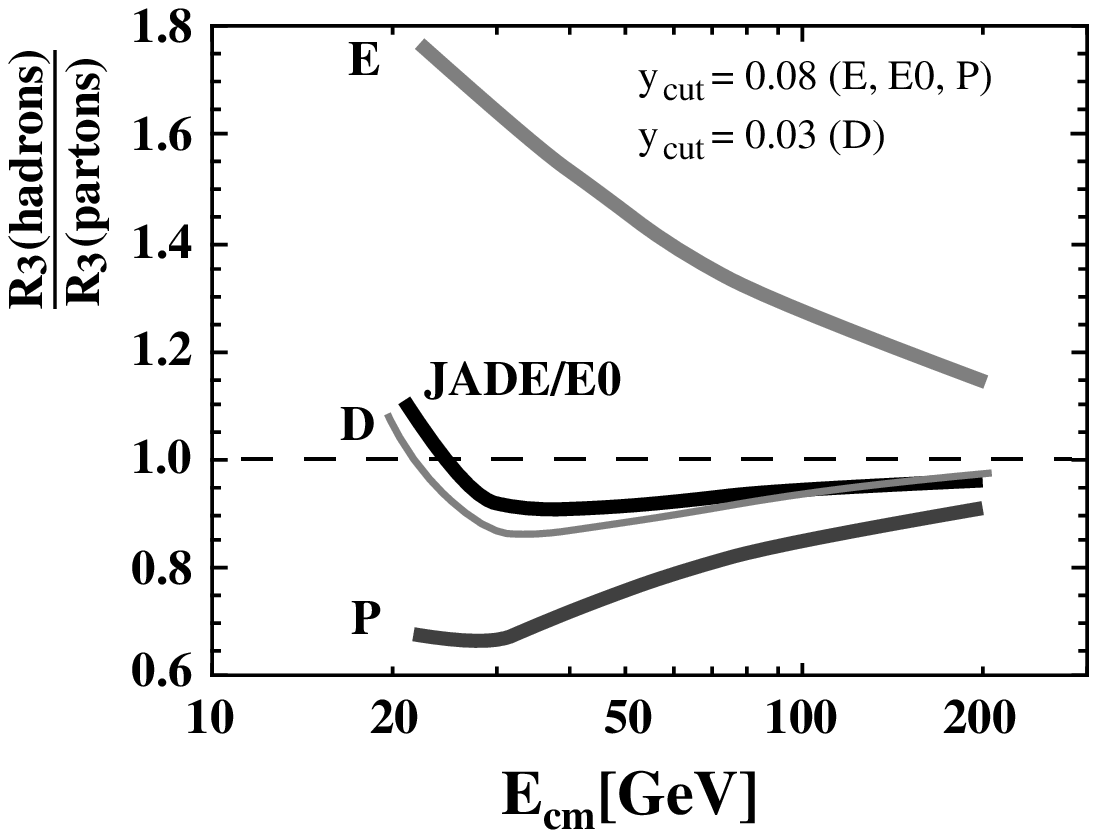}
\baselineskip=11.0pt
{\small      \noindent
{\bf Figure 1.}
The ratio $r$ of 3-jet event rates, calculated from JETSET QCD shower model
events before and after hadronisation, as a function of $\ecm$, for different
jet algorithms. }
\end{figure}
\baselineskip=12.0pt

\begin{figure}[htb]
\epsfxsize7.5cm\epsffile{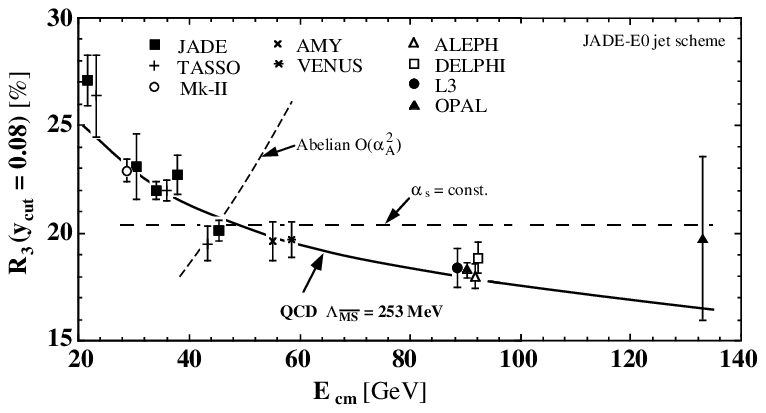}
\baselineskip=11.0pt
{\small      \noindent
{\bf Figure 2.}
Energy dependence of three-jet event production rates $R_3$, using the JADE E0
jet scheme with $\yc = 0.08$.
The measurements are compared with predictions of analytic $\oaa$ QCD
calculations, with the hypothesis of an energy independent $\as$ and with the
abelian vector theory in $\oabel$. }
\end{figure}
\baselineskip=12.0pt

Several methods of jet recombination and definitions of $M_{ij}$ exist, for which
QCD predictions in  complete $\oaa$ perturbation theory \cite{BKSS}, based on the
matrix elements of Ellis, Ross and
Terrano \cite{ert}, are available. 
In $\oaa$, the relative 2-, 3- and 4-jet production
rates, 
$R_n = \sigma_{\rm n-jet} / \sigma_{\rm tot}$, where $\sigma_{\rm tot}$ is
the total hadronic cross section and
$\sigma_{\rm n-jet}$ are the cross sections for $n$-parton event production, are
quadratic functions of the running coupling constant $\as (\mu)$.
In particular, 
    \begin{equation}
 R_3(y_c,\mu) = A(y_c)  \frac{\as(\mu)}{2\pi}  
   + B(y_c, x_{\mu})  \left(\frac{\as(\mu)}{2\pi}\right)^2 ,
      \end{equation}
\noindent
where $\mu = x_{\mu} \ecm$ 
is the renormalisation scale at which $\as$ is evaluated,
$x_{\mu}$ is the renormalisation scale factor
and $y_c \equiv \yc$.

The energy dependence of 
$R_3$ is only determined by the running $\as$; the scale factor $x_\mu$ 
- although it's optimal value is not given by the theory -
is not expected to change with energy.
Jet production rates are therefore
an ideal tool to test the energy dependence of $\as$, without the need
to actually determine $\as$ itself.

The first analysis in this sense was done by JADE \cite{jadejet2}.
The original JADE (also called $E0$) scheme with
$M_{ij}^2 = 2 E_i E_j (1-\cos{\theta_{ij}})$, where $E_i$ and $E_j$
are the energies
of the particles and $\theta_{ij}$ is the angle between them, has the
smallest hadronisation corrections with only a weak
dependence on the centre of mass energy, $\ecm$.
This is demonstrated in Figure~1, where the ratio $r$ =
$R_3$(hadrons) / $R_3$(partons) predicted by the JETSET QCD shower model
\cite{jetset} is plotted as a function of $\ecm$, for constant
values of $\yc$ \cite{scotland}.
For all jet algorithms, the quantity (1--$r$)
shows an approximate 1/$E_{cm}$ behaviour at large $E_{cm}$, as
expected for non-perturbative hadronisation effects. 
At smaller
energies, usually for $\sqrt{\yc} E_{cm} < 7$ GeV, $r$ increases with
decreasing $E_{cm}$ because of misassignments of jets, caused by hadronisation
fluctuations and heavy quark decays.
For the  JADE E0 algorithm, $|1 - r|$ is small enough and 
the energy dependence of $r$ is sufficiently
flat for $E_{cm}$ between 25 and 200 GeV to be
approximated by a constant within a systematic uncertainty
of $\pm 2\%$.
This feature makes
an important impact on the experimental evidence for
asymptotic freedom.
 
\begin{figure}[htb]
\epsfxsize7.5cm\epsffile{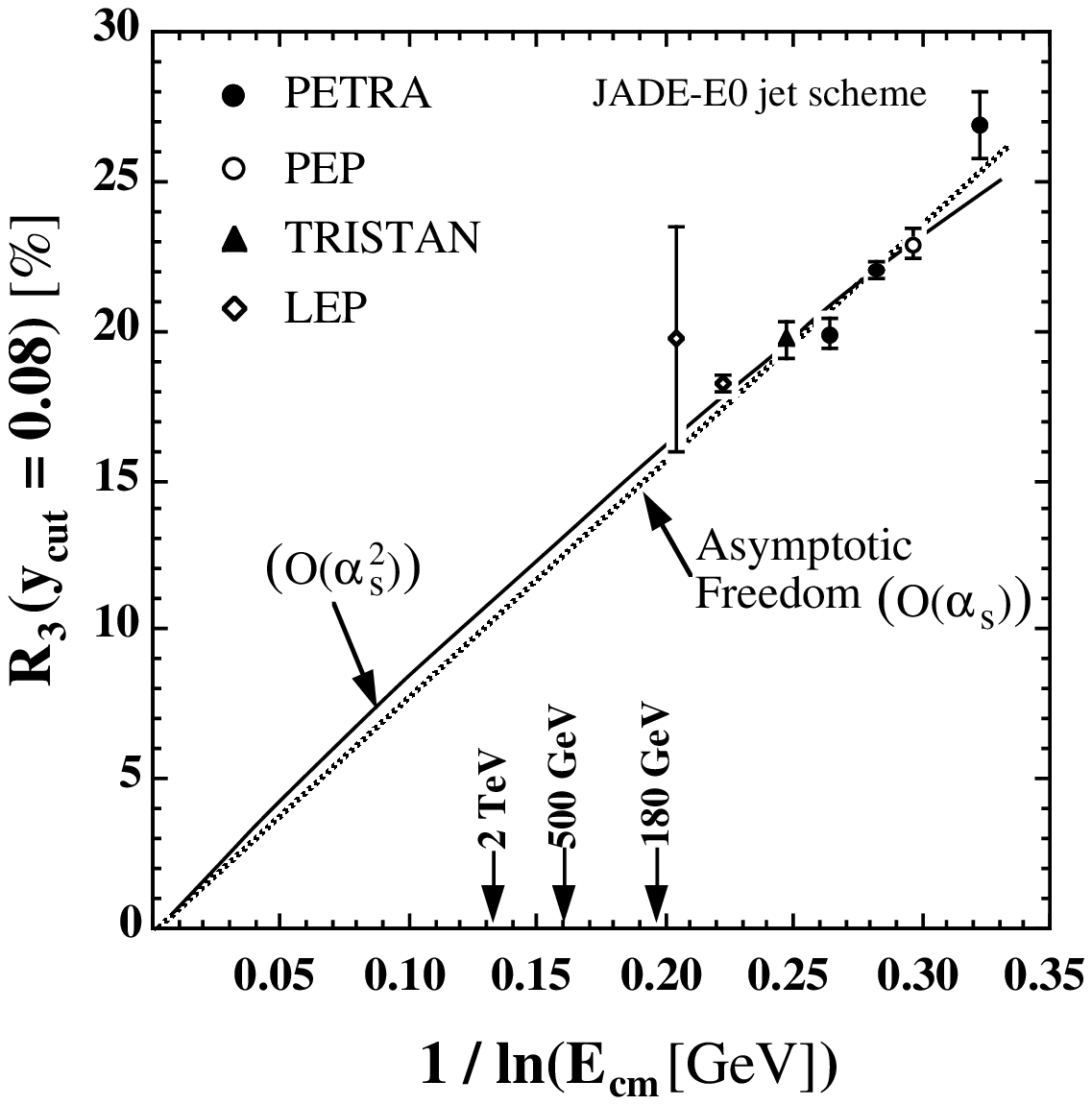}
\baselineskip=11.0pt
{\small      \noindent
{\bf Figure 3.}
The same data as shown in Fig.2, combined at similar energies, now as a
function of 1/ln($\ecm$). }
\end{figure}
\baselineskip=12.0pt

A compilation of the experimental results of $R_3$, analysed with
the JADE (E0) jet finder at different $\ecm$ using $\yc$ = 0.08,
is presented in Fig.~2 \cite{jadejet2,jetrates,o-133}.
The data are compared with fit results of analytic $\oaa$ QCD calculations
\cite{BKSS}, of the hypothesis of an energy $in$dependent coupling
constant and of the abelian, QED-like vector theory in $O(\alpha_A^2)$, where
$\alpha_A$ was adjusted such that the jet rates at $\ecm$ = 44 GeV are
reproduced \cite{scotland}.

For the QCD predictions and the hypothesis of $\as$ = constant,
the free parameters $\lamsb$ and a constant 3-jet rate $<R_3>$, 
respectively, were
determined by minimising $\chi^2$ for the data from $\ecm$ = 29 to 133 GeV.
The data points at $\ecm$ = 22 GeV were not included in the fit since
hadronisation effects may already bias the measurements at this energy, 
see Fig. 1.

For QCD, the fit results 
in $\lamsb = (253 \pm 12$~MeV), which corresponds to
$\amz = 0.120 \pm 0.001$ (stat. error only), and
$\chi^2  = 8.1$ for 13 degrees of freedom\footnote{
In order to account for the small, energy dependent hadronisation
effects as predicted by the model calculations shown in Fig. 1, 
a relative systematic point-to-point uncertainty of $\pm2\%$ is 
included when calculating $\chi^2$.},
corresponding to a confidence level (CL) of 84\%.
A linear fit through the data (not shown in Fig.~2) gives $\chi^2 = 12.7$ for 12
degrees of freedom (CL = 39\%).
The hypothesis of $\as$~= constant with
$\chi^2 = 72$ (CL = $3.4 \times 10^{-10}$) and 
the abelian theory, the $\chi^2$ of which tends to infinity, are entirely ruled
out by the data.

The experimental evidence for asymptotic freedom is further demonstrated in
Fig.~3, where the same experimental data, however combined at similar c.m.
energies, are plotted as a function of $1 / \ln (\ecm )$.
The dashed line is a fit to the leading order QCD prediction, namely $R_3
\propto \as \propto 1 / \ln \ecm$. 
The corresponding prediction in $\oaa$ is also shown, indicating that higher
order terms affect the energy dependence of $R_3$ only slightly.
At infinite energies $( 1 / \ln (\ecm ) \rightarrow 0 )$, $R_3$ and $\as$
are expected to vanish; an assumption which is in good agreement by the data.

While the most recent data from LEP-1.5 ($\ecm \sim 133$~GeV) are
statistically very limited, which will most likely not be improved
in the future, 
it is expected that LEP-2 ($\ecm \sim 175$~GeV) will provide another data point with an
absolute error of $\Delta(R_3) \sim 1 \%$.
The significance of data from future high energy $\epem$ linear
colliders can be inferred from  Figure~3: 
at $\ecm = 500$~GeV, the statistical error $\Delta(R_3)$
for 1000 hadronic events will
be about 1\%.
From the point of view of the previous experiments at PETRA and PEP and with the
eyes of QCD, i.e. on a logarithmic energy scale, a linear collider at $\ecm =
500$~GeV is almost half-way to infinte energies!
 
\section{RUNNING $\as$ FROM e-p COLLISIONS}

In deep inelastic electron-proton scattering (DIS),
hadronic final states can be studied in a wide range of energy scales $Q^2$,
the squared four-momentum transfer from the incident lepton.
Both experiments at HERA, H1 and ZEUS, determined $\asq$ from jet rates
measured in the energy range of $10~{\rm GeV}^2 < Q^2 < 4000~{\rm GeV}^2$
\cite{h1jet,zeusjet}, using a modified
JADE jet algorithm where the proton remnant is treated as a pseudoparticle
which carries only longitudinal momentum (i.e. along the beam direction).

The jet resolution parameter is given by $y_c = M_{ij}^2 / W^2$, where
$W$ is the invariant mass of the hadronic system and $M_{ij}$  the invariant
pair masses of all objects used including the pseudoparticle.
This choice of algorithm ensures a jet classification which is similar to
that used in $\epem$ annihilation. 
The jet production rates $R_{N+1}$ are calculated from the number of events
with $N$ resolvable jets inside the acceptance region, where \oq +1" denotes
the pseudoparticle (or pseudojet) from the proton remnant.

In DIS, jet rates by themselves cannot demonstrate the running of the coupling
since theory predicts that the QCD coefficients of $\as$ also depend, in
contrast to the case of $\epem$ annihilation (see Eq.~3), on $Q$ through the
parton density functions.
QCD predictions for (1+1) and (2+1)-jet events which are complete to
$\oaa$ \cite{graudenz} are therefore used to extract $\as$ for
different bins of $Q^2$.

\begin{figure}[tb]
\epsfxsize7.5cm\epsffile{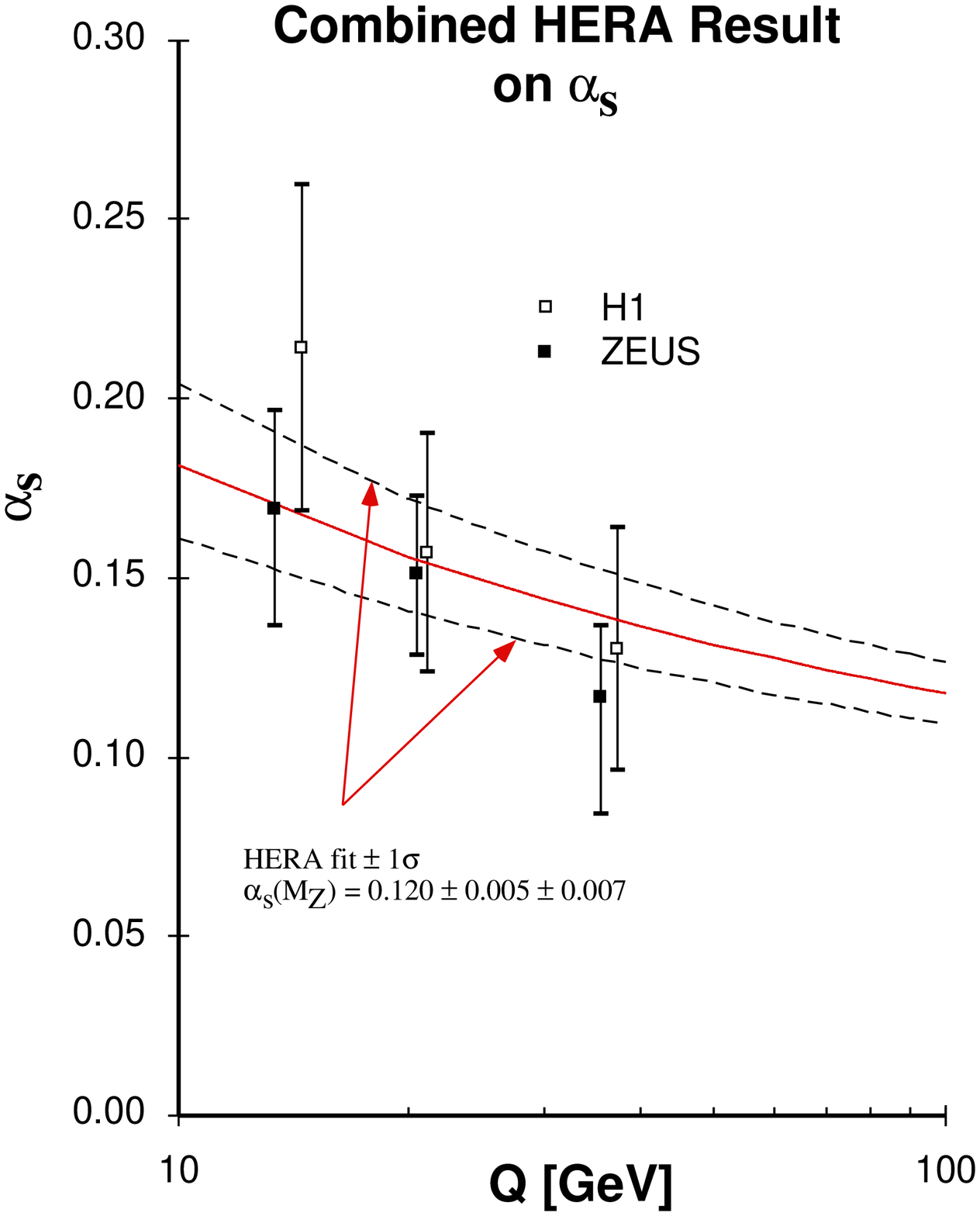}
\baselineskip=11.0pt
{\small      \noindent
{\bf Figure 4.}
Measurements of $\asq$ from jet rates at HERA. The curves are the results of a
QCD fit to the data (compilation from \cite{wwwh1}). }
\end{figure}
\baselineskip=12.0pt
 
The results are compiled in Fig.~4 \cite{wwwh1}.
Although the overall uncertainties, both statistical as well as systematic, are
still rather large in these measurements, the general trend of a coupling which
decreases with increasing $Q$ can clearly be seen.
The lines in Fig.~4 indicate the results of a QCD fit through the measurements,
which extrapolates to $\amz = 0.120 \pm 0.005 \pm 0.007$.

In another study of the HERA data \cite{ball}, $\as$ is determined from
the proton structure function ${\rm F}^p_2 (x,Q^2)$ at small $x$ and $Q^2 <
100~{\rm GeV}^2$.
$F^p_2$ is computed in next-to-leading order in $\as$, including summations
of all leading and subleading logarithms of $Q^2$ and $1/x$.
In that study it is
demonstrated that the structure function data of H1 and of ZEUS exhibit double
logarithmic scaling in both $x$ and $Q^2$, which is regarded as direct evidence
for the running $\as$
\cite{ball}. 
A QCD fit to these data finally gives $\amz = 0.120 \pm 0.005 \pm
0.009$, where the first error is experimental and the second theoretical.
This is in good agreement with the result from jets described above, and
also with the world average of $\amz$, see section~7.
 
\section{JETS IN HADRON COLLISIONS}

Similarly as in deep inelastic lepton-nucleon scattering, hadron colliders
provide the opportunity to simultaneously probe QCD in a wide range of
momentum transfers $Q$.
In a recent study \cite{gielejets} based on the one-jet inclusive transverse
energy ($E_T$) distribution measured at the Tevatron \cite{cdfjets},
values of
$\as (Q \equiv E_T)$ are determined over a wide range of energies, $E_T = 30\
{\rm to}\ 500$~GeV.

\begin{figure}[tb]
\epsfxsize7.5cm\epsffile{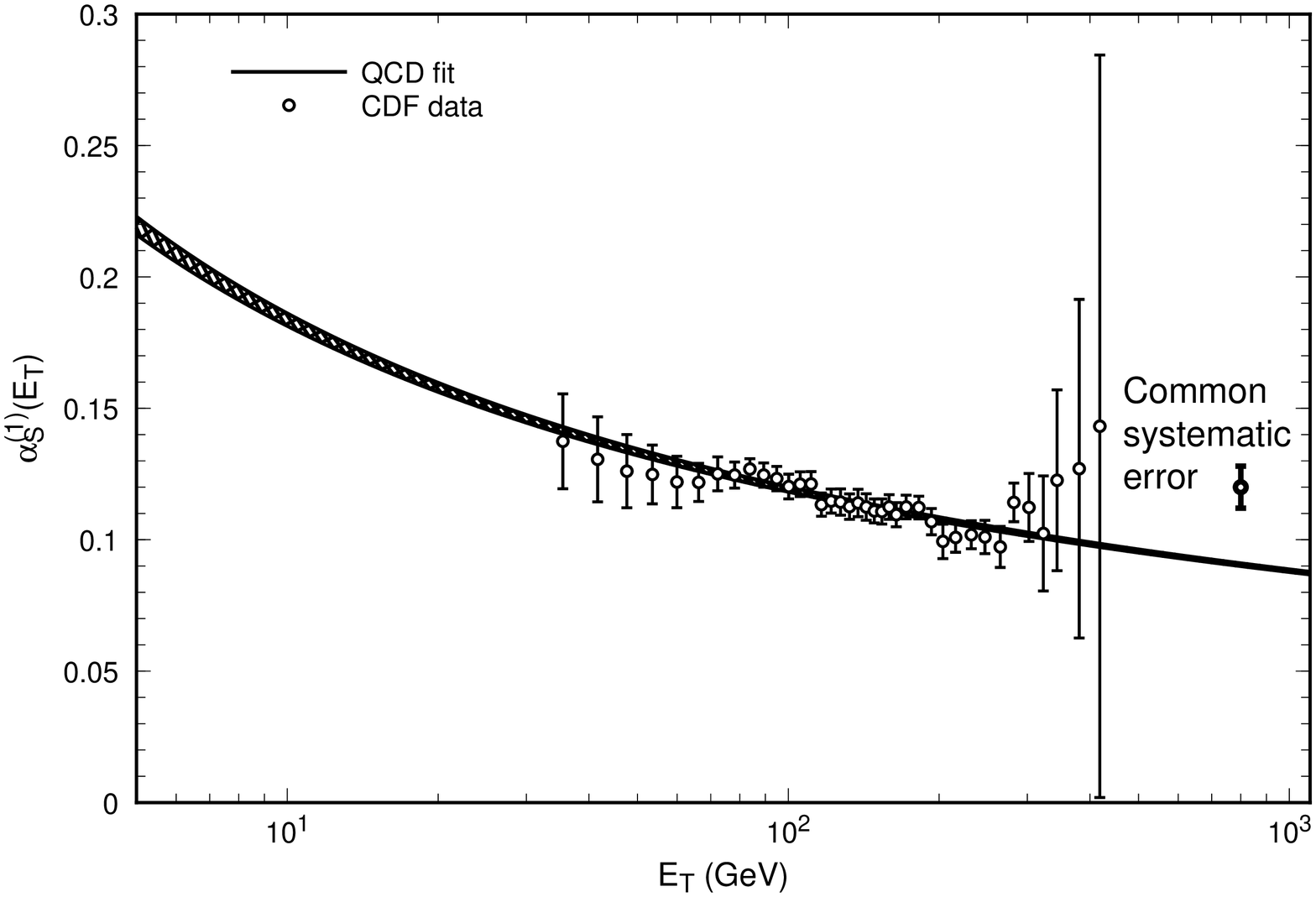}
\baselineskip=11.0pt
{\small      \noindent
{\bf Figure 5.}
Values of $\as (E_T)$ extracted from the one-jet inclusive jet cross sections
from CDF as a function of the the jet transverse energy $E_T$, together with
the QCD expectations based on NLO perturbation theory and the MRSA' particle
density function (from \cite{gielejets}). }
\end{figure}
\baselineskip=12.0pt

The results of this study are shown in Fig.~5.
The values of $\as (E_T)$ are seen to decrease with increasing $E_T$, in good
agreement with the QCD expectations of a running coupling strength 
(shaded area). 
A simultaneous QCD fit to these data results in $\amz = 0.121 \pm 0.001\ {\rm
(stat.)}\
\pm 0.008\ {\rm (syst.)}\ \pm 0.005\ {\rm (theor.)}$, which is in excellent
agreement with other measurements, see section~7.

\section{RUNNING $\as$ FROM $\tau$ DECAYS}

Measurements of the ratio of the hadronic and leptonic branching
fractions of the $\tau$ lepton, $R_{\tau}$, have provided precise values of
$\as$ at the energy scale of the $\tau$-mass, $Q \equiv M_{\tau} = 1.777\
{\rm GeV}$, see e.g.
\cite{qcd94,duflot} and references quoted therein.
Recently, a new test of the energy dependence of $\as$ was proposed
\cite{neubert}, based on
the $\tau$ decay rate into hadrons of invariant mass squared $s$ smaller than
a threshold value $s_0$:
\begin{eqnarray}
R_{\tau} (s_0) &=& \frac{\Gamma (\tau \rightarrow \nu_{\tau} + {\rm hadrons};\ 
s < s_0)}{\Gamma (\tau \rightarrow \nu_{\tau} e \overline{\nu_e})}
\nonumber \\
&=& \int^{s_0}_0 {\rm d} s \frac{{\rm d}R_{\tau} (s)}{{\rm d}s}.
\end{eqnarray}
The running coupling constant $\as (s_0)$ is extracted 
from the inclusive hadronic spectrum ${\rm d}R_{\tau} (s) / {\rm d}s$
measured by ALEPH and CLEO
\cite{duflot,cleotau}, in the low energy region $0.7\ {\rm GeV}^2 < s_0 <
M_{\tau}^2$ where $\as$ is expected to change by almost a factor of two.

\begin{figure}[tb]
\epsfxsize7.5cm\epsffile{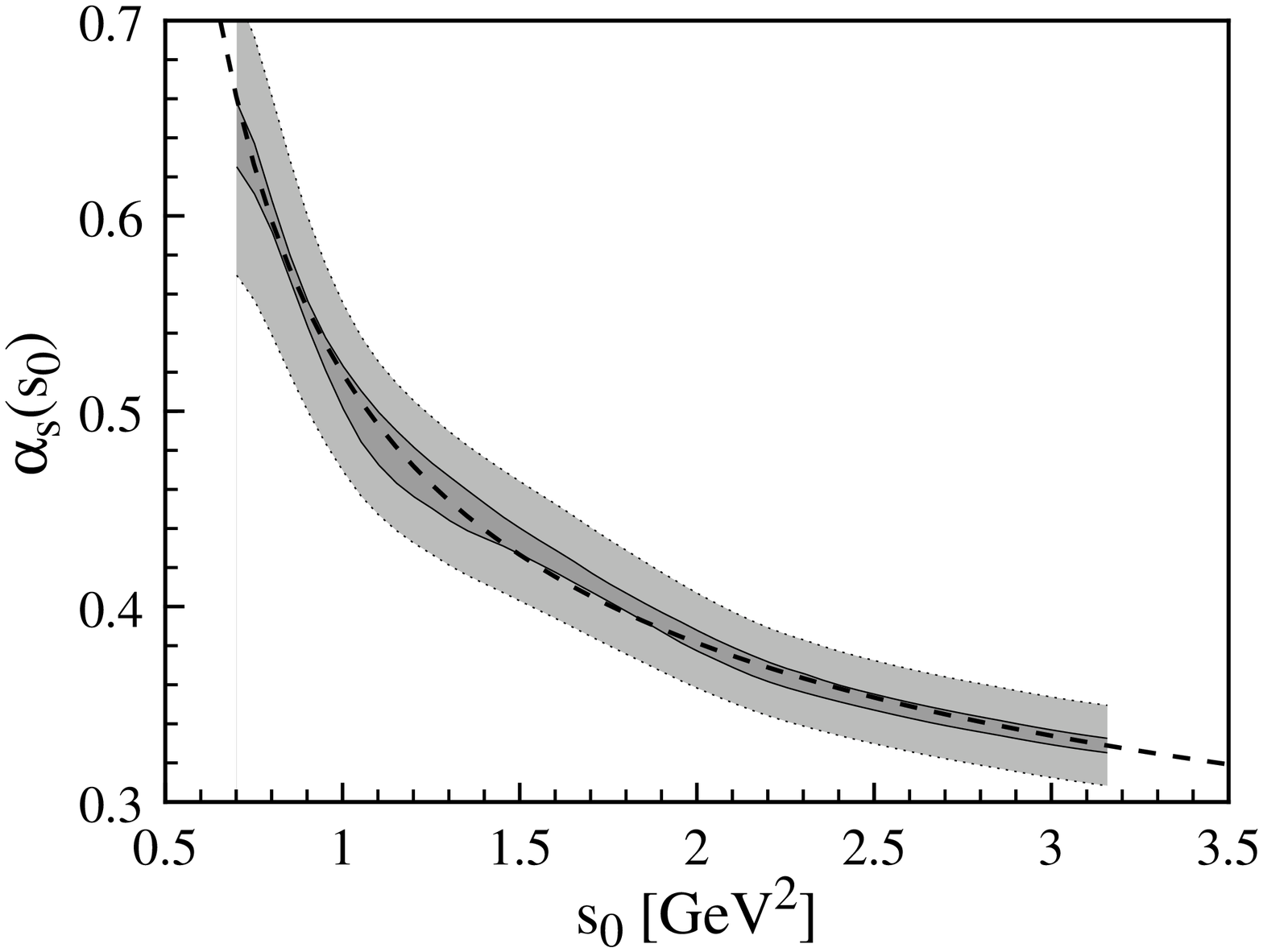}
\baselineskip=11.0pt
{\small      \noindent
{\bf Figure 6.}
Values of $\as (s_0)$ from the data on $R_{\tau} (s_0)$. 
The inner band represents experimental, the outer band the sum of experimental
and theoretical uncertainties.
The dashed line shows the running coupling constant in $\oaaa$ QCD.
(From \cite{neubert}) }
\end{figure}
\baselineskip=12.0pt

Theoretical predictions for $R_{\tau} (s_0)$ include perturbative terms
which are complete to 
$\oaaa$  as well as estimates of nonperturbative
contributions using the operator product expansion \cite{shifman,braaten}.
Assuming global parton-hadron duality, $\as$ can thus be determined from each
measured value of $R_{\tau} (s_0)$.

The results of the study of ref.\cite{neubert} are shown in Fig.~6.
Since values of $\as$ extracted from $R_{\tau} (s_0)$ are correlated with each
other, the fit results are displayed as a band.
The dashed curve  shows the QCD expectation of the running coupling
constant calculated in $\oaaa$,
normalised to the data at $s_0 = M^2_{\tau}$.

The observed energy dependence is in excellent agreement with the QCD
prediction of the running $\as$ in $\oaaa$.
In addition, the data show a distinct preference for the 3-loop
$\beta$-function (Eq.~ 2), compared to the leading order (1-loop) one 
\cite{neubert}.
The overall value for $\as$, including estimates of higher order perturbative
uncertainties, results in $\as (M_{\tau}) = 0.33 \pm 0.03$ or, equivalently,
in $\amz = 0.119 \pm 0.004$.

%
\begin{table*}[htb]
\begin{center}
\begin{tabular}{|l|c|c|l|l|c c|c|}
   \hline
 & &  Q & & &  \multicolumn{2}{c|}
{$\Delta \amz $} &  \\ 
 Process & Ref. & [GeV] & $\alpha_s(Q)$ &
  $ \amz$ & exp. & theor. & Theory \\
\hline \hline \normalsize
 & & & & & & & \\
 DIS [$\nu$; Bj-SR] & \cite{bj-sr-ellis} & 1.58
  & $0.375\ ^{+\ 0.062}_{-\ 0.081}$ & $0.122\ ^{+\ 0.005}_{-\ 0.009}$ & 
  -- & -- & NNLO \\
 DIS [$\nu$; GLS-SR] & \cite{gls-theory} & 1.73
  & $0.32\pm 0.05$ & $0.115\pm 0.006$ & $ 0.005 $ & $ 0.003$ & NNLO \\
 & & & & & & & \\
 $\tau$-decays & \cite{narison-i1,neubert}
  & 1.78 & $0.330 \pm 0.030$ & $0.119 \pm 0.004$
  & 0.001 &  0.004 & NNLO \\
 & & & & & & & \\
 DIS [$\nu$; ${\rm F_2\ and\ F_3}$] & \cite{ccfr} & 5.0
  & $0.193\ ^{+\ 0.019\ }_{-\ 0.018\ }$
   & $0.111\pm 0.006$   &
    $ 0.004 $ & $ 0.004$ & NLO \\
 DIS [$\mu$; ${\rm F_2}$]& \cite{virchaux}
     & 7.1 & $0.180 \pm 0.014$ & $0.113 \pm 0.005$ & $ 0.003$ &
     $ 0.004$ & NLO \\
 DIS [HERA; jets] & \cite{h1jet,zeusjet}
     & 10 - 60 &  & $0.120 \pm 0.009$ & $ 0.005$ &
     $ 0.007$ & NLO \\
 DIS [HERA; ${\rm F_2}$] & \cite{ball}
     & 2 - 10 &  & $0.120 \pm 0.010$ & $ 0.005$ &
     $ 0.009$ & NLO \\
& & & & & & & \\
 ${\rm Q\overline{Q}}$ states & \cite{davies}
     & 5.0 & $0.203 \pm 0.007$ & $0.115 \pm 0.002 $ & 0.000 & 0.002
     & LGT \\
 $J/\Psi + \Upsilon$ decays & \cite{kobel}
     & 10.0 & $0.167\ ^{+\ 0.015\ }_{-\ 0.011\ }$ & $0.113\ ^{+\ 0.007\ }
     _{-\ 0.005\ }$ & 0.001 & $^{+\ 0.007}_{-\ 0.005}$ & NLO \\
 & & & & & & & \\
 $\epem$ [$\sigma_{\rm had}$] & \cite{haidt} & 34.0 &
 $0.146\ ^{+\ 0.031}_{-\ 0.026}$ &
   $0.124\ ^{+\ 0.021}_{-\ 0.019}$ & $^{+\ 0.021}_{-\ 0.019}
   $ & -- & NLO \\
 $\epem$ [ev. shapes] & \cite{budapest} & 35.0 & \ $0.14\pm 0.02$ &
   $0.119 \pm 0.014$ & -- & -- & NLO \\
 $\epem$ [ev. shapes]  & \cite{topazas}  & 58.0 & $0.132\pm 0.008$ &
   $0.123 \pm 0.007$ & 0.003 & 0.007 & resum. \\
 & & & & & & & \\
 $p\bar{p} \rightarrow {\rm b\bar{b}X}$  & \cite{ua1-bb}
    & 20.0 & $0.145\ ^{+\ 0.018\ }_{-\ 0.019\ }$ & $0.113 \pm 0.011$ 
    & $^{+\ 0.007}_{-\ 0.006}$ & $^{+\ 0.008}_{-\ 0.009}$ & NLO \\
 ${\rm p\bar{p},\ pp \rightarrow \gamma X}$  & \cite{a-ua6} & 24.2 & $0.137
 \ ^{+\ 0.017}_{-\ 0.014}$ &
  $0.112\ ^{+\ 0.012\ }_{-\ 0.008\ }$ & 0.006 &
  $^{+\ 0.010}_{-\ 0.005}$ & NLO \\
 ${\sigma (\rm p\bar{p} \rightarrow\  jets)}$  & \cite{gielejets} & 30 - 500 & 
&
  $0.121\pm 0.009$ & 0.001 & 0.009 & NLO \\
 & & & & & & & \\
$\epem \rightarrow \z0$:  & & & & & & & \\
 \ \ $\Gamma (\z0 \rightarrow {\rm had.})$ & \cite{lep-ewwg}
    & 91.2 & $0.126\pm 0.006$ & 
$0.126\pm 0.006$ &
   $ 0.005$ & $0.003$ & NNLO \\
 \ \ had. event shapes & \cite{qcd94} &
    91.2 & $0.119 \pm 0.006$ & $0.119 \pm 0.006$ &$ 0.001$ & $ 0.006$
& NLO\\ 
 \ \ had. event shapes & \cite{qcd94} &
    91.2 & $0.122 \pm 0.006$ & $0.122 \pm 0.006$ & $ 0.001$ & $
0.006$ & resum. \\
 & & & & & & & \\
$\epem$ [ev. shapes]  & Tab. 2 & 133.0 & $0.112\pm 0.009$ &
   $0.118 \pm 0.009$ & 0.003 & 0.009 & resum. \\
 & & & & & & & \\
\hline
\end{tabular}
\end{center}
\baselineskip=11.0pt
{\small      \noindent 
{\bf Table 1.}
World summary of measurements of $\as$.
Abbreviations:
DIS = deep inelastic scattering; GLS-SR = Gross-Llewellyn-Smith sum rules;
Bj-SR = Bjorken sum rules;
(N)NLO = (next-)next-to-leading order perturbation theory;
LGT = lattice gauge theory;
resum. = resummed next-to-leading order.}
\end{table*}
%
\section{WOLRD SUMMARY OF $\as$}

Having discussed the current tests of asymptotic freedom from measurements
based on single observables like jet rates, jet-$E_T$-spectra and
$\tau$-decays, the overall summary of all available $\as$ determinations
remains to be updated (see e.g. \cite{altarelli,sb-catani,qcd94,moriond95} for
previous reviews).

An update of the summary given in \cite{qcd94} is presented in
Table~1.
Graphical presentations of the running coupling $\as (Q)$ and of the results 
extrapolated to $\amz$, using the 3-loop expansion (Eq.~2) and treating
heavy flavour thresholds according to ref. \cite{wernerb}, are given in
Figs.~7 and~8, respectively.
The most recent changes and additions are discussed in the
following subsections; see \cite{qcd94} for comparison.

\subsection{$\as$ from Sum Rules}
The results from the Gross-Llewellyn-Smith and the Bjorken sum rules
\cite{bj-sr-ellis,gls-theory} have been retained; a preliminary update of the
Bjorken sum rule result from the CCFR collaboration (see e.g.
\cite{harris-m95}) exists but is not included here because the final
publication is still missing.

\subsection{$\as$ from $\tau$ Decays}
In the previous report \cite{qcd94}, $\as$ from $\tau$ decays was obtained from
a compilation of measurements of the ratio of the hadronic to the leptonic
$\tau$ branching ratios, $R_\tau$.
Meanwhile, several new measurements of this quantity became available (see
\cite{duflot,opaltau} and references quoted therein).
Instead of deriving an updated value of $\as$ from $R_\tau$, the
results from the study described in section~6, namely from the hadronic
invariant mass distribution of $\tau$ decays, $R_{\tau} (s_0)$, is taken.
This result is identical to the one derived from a recent evaluation of $\as$
and its overall uncertainty from $R_\tau$ \cite{narison-i1}:
$\as (M_\tau) = 0.33 \pm 0.03$.

\subsection{$\as$ from Deep Inelastic Scattering}
In addition to the earlier results from fixed target experiments
\cite{ccfr,virchaux}, 
the new values of $\as$ from HERA, discussed in Sect.~4, are included.

\begin{figure}[tb]
\epsfxsize7.5cm\epsffile{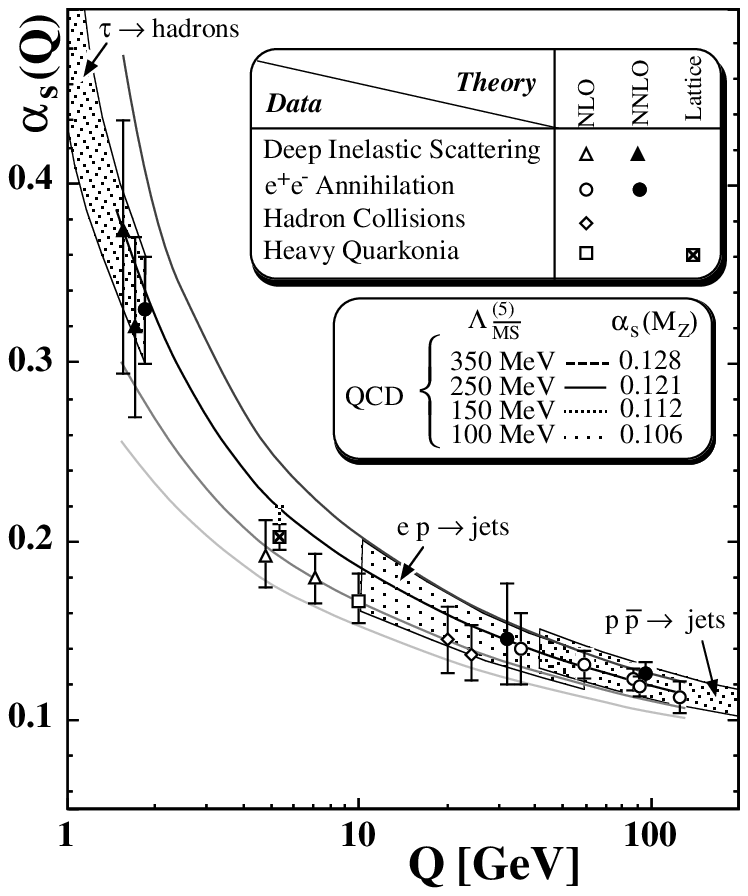}
\baselineskip=11.0pt
{\small      \noindent
{\bf Figure 7.}
A Summary of measurements of $\as$, compared with QCD
expectations for four different values of
$\lamsb$ which are given for $N_f = 5$ quark flavours
(relation between $\as$ and $\lamsb$ in $\oaaa$).
}
\end{figure}
\baselineskip=12.0pt
 
\begin{figure}[tb]
\epsfxsize7.5cm\epsffile{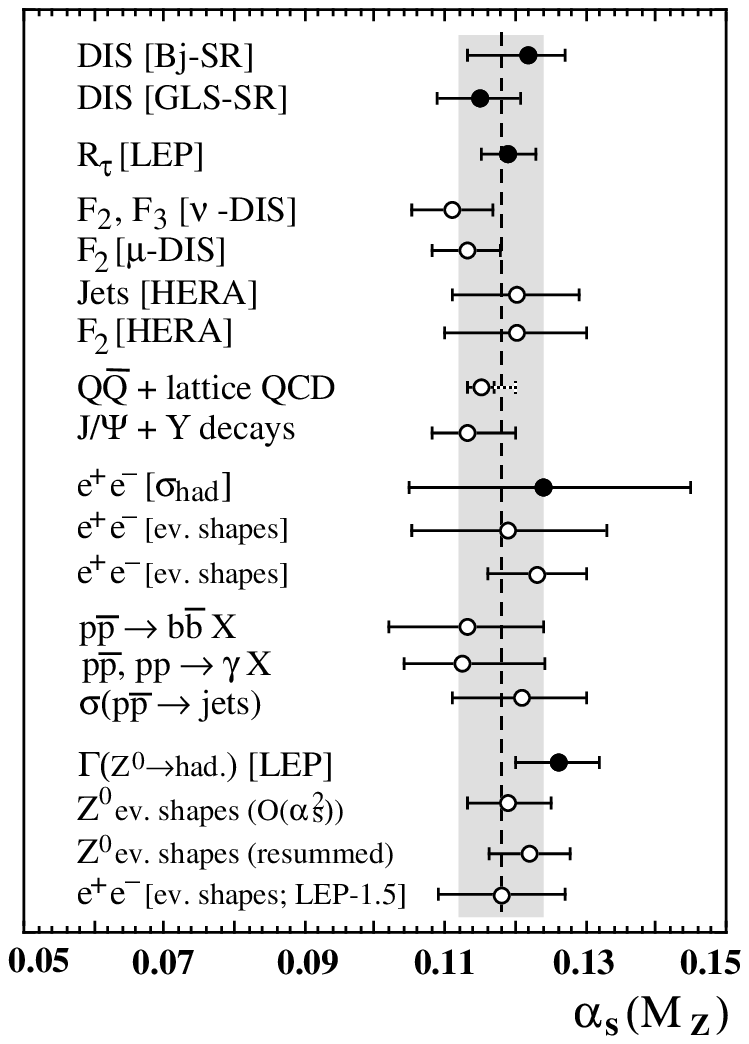}
\baselineskip=11.0pt
{\small      \noindent
{\bf Figure 8.}
A Summary of measurements of $\amz$, as listed in Table~1.
Filled symbols are derived using $\oaaa$ QCD; open symbols are in $\oaa$ or
based  on lattice calculations.}
\end{figure}
\baselineskip=12.0pt

\subsection{$\as$ from Lattice QCD}
The values of $\amz$ from lattice QCD calculations, based on measurements of
heavy quarkonia mass spectra, slowly but gradually increased during the past
few years.
These changes are mainly due to the availability of unquenched calculations
(i.e. including dynamical light quark flavours) and to more refined procedures
to convert the lattice coupling to the running coupling of perturbative QCD.
A recent summary of these results gives $\amz = 0.115 \pm 0.002$ \cite{davies},
which is taken over for this review.
There are, however, unpublished reports which result in 
$\amz = 0.118 \pm 0.002$, see e.g. \cite{lepage}.

\subsection{$\as$ from Hadron Collsions}
The previous, preliminary determination of $\as$ from a measurement of the 
$b \overline{b}$ cross section was updated and finally published in
\cite{ua1-bb}.

Results on $\as$ from ${\rm p\bar{p} \rightarrow W\ jets}$ 
\cite{a-ua2,a-ua1}  which were
considered in previous compilations are no longer included since the QCD
calculations on which they are based are not complete to next-to-leading order.
A recent study of this process from D0 reports that calculations which are 
complete to NLO do not provide a reasonable fit of the data \cite{a-d0}.

The result on $\as$ from the one-jet inclusive $E_t$-distribution, as
discussed in Section~5, is a new entry in Table~1.

\subsection{$\as$ from the $\z0$ Line Shape}
The value of $\amz$ derived from the hadronic width of the $\z0$ boson was
continuously updated during the past years, according to the increasing data
statistics of the four LEP experiments.
Not all of these updates were published in journals, however they
are documented as CERN preprints which are commonly available.
In this review, the result which was documented last before this conference
is taken,
$\amz = 0.126 \pm 0.006$.
\cite{lep-ewwg}.

\subsection{$\as$ from Event Shapes at LEP-1.5}
Three of the LEP experiments have published determinations of $\as$ from the
data taken at $\epem$ c.m. energies between 130 and 136~GeV 
\cite{l-133,a-133,o-133}.
Each experiment collected about 5~$pb^{-1}$ of data, corresponding to only
about 300 non-radiative hadronic events per experiment.
Due to the large statistical uncertainty of each experiment, only the $combined$
value of 
$\as$ from LEP-1.5 can provide a  meaningful test of the running coupling.

The results of $\as$ from jet rates and from hadronic event shapes at
LEP-1.5 are summarised in Table~2. 
There is good agreement between the experiments, within the statistical
uncertainties. The average result is
$\as (133\ {\rm GeV}) = 0.112
\pm 0.009$ or, equivalently,
$\amz = 0.118 \pm 0.009$, which is compatible, within the experimental errors,
with the value which was directly obtained at the $\z0$ resonance,
$\amz = 0.121 \pm 0.006$.

\subsection{World Average of $\amz$}

Averaging the values of $\amz$ from Table~1, either unweigthed or weigthed by
the inverse square of their errors, gives
\footnote{The result which is based on lattice QCD is not included when
computing the weigthed average; see Section~7.4 for
justification.} 
$\overline{\as} (\mz ) = 0.118$ in both cases.
This value has been remarkably stable during the past few years, see e.g. 
\cite{altarelli,sb-catani,qcd94,moriond95} for
previous reviews.
From Fig.~8 it can be seen that all results of $\as$ are compatible with
this world average, within the errors assigned to the measurements.

The errors of most $\as$ results are dominated by theoretical uncertainties,
which are estimated using a variety of different methods and definitions.
The significance of the quoted errors is largely unknown; they are neither
gaussian nor are the correlations between different
measurements known.
A \oq correct" calculation of the overall uncertainty of
$\overline{\as} (\mz )$ is therefore not possible. 

Some methods were proposed to compute the overall error from the individual
ones, either by rescaling the latter or by constructing an ad-hoc correlation
matrix such that the overall $\chi^2$ deviation from the mean value is equal
to the number of degrees of freedom (i.e. to $n-1$, where $n$ is the number of
individual measurements) \cite{pdg96,schmelling}.
If applied to the results listed in Table~1, these methods suggest that
$\Delta \overline{\as}(\mz) \sim 0.003 ... 0.005$.

Since most of the errors listed in Table~1 are not gaussian but rather
indicate probability distributions of rectangular shape
(however still with unknown correlations between each other), 
a more pragmatic
and conservative estimate of the overall uncertainty of $\overline{\as}(\mz)$ is
therefore applied:
counting the relative number of entries in Table~1 whose central values are
within $\pm \Delta \overline{\as}$ of
$\overline{\as} (\mz ) = 0.118$, one gets about
45\% for $\Delta \overline{\as}= 0.003$, 60\% for 0.004, 75\% for 0.005,
90\% for 0.006, 95\% for 0.008 and 100\% for 0.008.
A 90\% \oq confidence level" seems to be a reasonable and safe estimate for
$\Delta \overline{\as}$, such that the world
average is quoted to be
$$ \overline{\as}(\mz) = 0.118 \pm 0.006\ ,$$
which corresponds, in $\oaaa$ and for $N_f = 5$ or 4 flavours, to
$$ \Lambda^{(5)}_{\overline{MS}} = 210^{+80}_{-65}\ {\rm MeV,\ or\ \ }
   \Lambda^{(4)}_{\overline{MS}} = 295^{+95}_{-80}\ {\rm MeV.} $$
The world average 
is indicated by the vertical line and the shaded area in Fig.~8.

\begin{table*}[htb]
\begin{center}
\begin{tabular}{|l|c|c|c||c|c|}
   \hline
Exp. & Ref. & $\as$(133 GeV) & $\rightarrow \amz$ [LEP-1.5]& 
$\amz$ [LEP-I] & $\#\sigma$ \\
\hline
ALEPH & \cite{a-133} & $0.119 \pm 0.005 \pm 0.007$ &
$0.126 \pm 0.006 \pm 0.008$ & $0.120 \pm 0.002 \pm 0.007$ & 0.95 \\
L3 & \cite{l-133} & $0.107 \pm 0.005 \pm 0.006$ & $ 0.113 \pm
0.006 \pm 0.007$ & $ 0.125 \pm 0.003 \pm 0.008$ & 1.8 \\
OPAL & \cite{o-133} & $0.110 \pm 0.005 \pm 0.009$ &
$ 0.116 \pm 0.006 \pm 0.010 $ & $0.120 \pm 0.002 \pm 0.006$& 0.63 \\
\hline
Average & & $0.112 \pm 0.003 \pm 0.008$ & $ 0.118 \pm 0.003 \pm 0.009$ &
$ 0.122 \pm 0.001 \pm 0.006$ & 0.95 \\
\hline
\end{tabular}
\end{center}
\baselineskip=11.0pt
{\small      \noindent 
{\bf Table 2.}
Summary of measurements of $\as$ at LEP-1.5 ($<\ecm >$ = 133 GeV).
The first errors are experimental, the second theoretical.
The last two columns give the results of $\amz$ previously obtained at the
$\z0$ resonance (LEP-I) and the number of standard deviations between
the LEP-I and LEP-1.5 results, respectively, taking only experimental errors
into account.}
\end{table*}
%

\subsection{Systematic Differences in $\amz$?}

In previous reviews the observation was made
that measurements which are obtained at energy scales of 5~GeV~$< Q
<$~20~GeV are systematically low, corresponding to $\amz \approx 0.112$, while
at $Q \approx M_\tau$ and $Q \ge 30$~GeV
the results tend to be higher, $\amz \approx 0.120$ and 0.122, respectively.
Speculations about the origin of these differences include the
existence of a light, neutral, coloured object of spin 1/2 (e.g. a
gluino), a possible dependence on the scattering process ($\epem$
annihilation or deep inelastic scattering) or effects of quark
masses which are not included in the current higher order QCD calculations
\cite{qcd94}.

In general, these systematic but hardly conclusive differences are still visible
in this summary, see Table~1 and Figure~8. 
However, the most recent results of $\as$ from deep inelastic
scattering at HERA and from jet production in hadron collisions underline the
tendency towards higher values of $\amz$, in agreement with those from $\epem$
annihilation, and the detailed studies of hadronic $\tau$-decays provide
consistent results, too.
Therefore the hyptheses of a process dependence of $\as$ or of the existence of
light gluinos are not very likely to explain the suspected differences.
The most probable origin of those, if significant at all, could be the absence
of heavy quark mass effects in the current QCD calculations, which would
affect the results from data with energies close to the quark thresholds most.

Next-to-leading order calculations including quark mass effects are currently
being worked on.
This, together with the ongoing efforts to determine $\as$ from the yet
increasing amount of data from various processes, has the potential to
decrease systematic uncertainties and to resolve the cause of the differences
which are still being observed.
 
\section{Conclusion}

Asymptotic freedom, which is $the$ key feature of
the theory of strong interactions, has been successfully tested in various
experimental studies.
Perhaps the most intuitive and direct method, the study of the energy
dependence of 3-jet event production rates in
$\epem$ annihilation, $R_3$, began to provide evidence
for the running of $\as$ already in 1988.
These studies, carried out by various experiments in a large range of c.m.
energies, are based on
the JADE-E0 jet algorithm for which,
at constant jet resolution, hadronisation corrections are small
enough such that $R_3$ is directly proportional to $\as$. 
With the availability of the LEP data the evidence developed into a
proof of asymptotic freedom, demonstrating that $\as$ decreases
with increasing energy, as predicted by QCD.

Further significant tests became available in the past two years:
jet rates in ep-collisions at HERA and 
1-jet inclusive transverse energy distributions in hadron collisions,
measured by single experiments in large regions of the energy scale,
provide the possibility to determine the energy dependence of $\as$ 
while minimising systematic point-to-point uncertainties.
A new analysis based on invariant mass distributions of hadronic
$\tau$-decays demonstrates the running of $\as$ in the energy
range of $0.7~{\rm GeV}^2 < Q^2 < M^2_\tau$, where the coupling changes by
almost a factor of two.

In addition to these dedicated tests, the world summary
of measurements of $\as$, in the energy range of $M_\tau \le Q \le 133$~GeV, 
provides compelling evidence for asymptotic freedom.
Within their assigned uncertainties,
all measurements are compatible with the QCD expectation of a running $\as$.

Extrapolated to a common energy scale, using Equations 1~and~2 and treating
quark flavour thresholds as described in \cite{wernerb,pdg96}, the measurements
of $\as$ average to
$$\amz = 0.118 \pm 0.006\ .$$
The overall uncertainty of 0.006 corresponds to a simple estimate
of a \oq 90\% confidence level", derived from the scatter of the individual
results.

\bigskip
\noindent {\bf Acknowledgements.}
It is a pleasure to thank S. Narison for providing the possibility to 
present this review at this well organised and informative conference.
I am grateful to E. Elsen, W. Giele, and M. Neubert for providing 
and authorising the use of Figures 4, 5 and 6, and to W. Bernreuther
for many interesting discussions.


\end{document}